\documentclass[apsrev4-2,aps,physrev,superscriptaddress,twocolumn,floatfix,longbibliography]{revtex4-2}
\usepackage{colortbl}
\usepackage{color}
\usepackage{comment}
\usepackage{amsfonts,amsmath,amssymb}
\usepackage[dvipsnames]{xcolor}
\usepackage{tabularx,hhline,tikz,multirow,array}
\usepackage{bm,enumitem}
\usepackage{dcolumn,stmaryrd}
\usepackage[colorlinks=true,pdfborder={0 0 0},linkcolor=red,citecolor = blue]{hyperref}
\usepackage[normalem]{ulem}

\newcommand{\Sp}{\mathbf{S}}

\def\rv {{\bf r}}
\def\kv {{\bf k}}
\def\qv {{\bf q}}

\allowdisplaybreaks
\begin{document}
\title{Frustration-driven topological textures on the honeycomb lattice: antiferromagnetic meron-antimeron and skyrmion crystals emerging from spiral spin liquids}

\author{M. Mohylna}
\email{mohylna@kth.se}
\affiliation{Department of Applied Physics, School of Engineering Sciences, KTH Royal Institute of Technology, AlbaNova University Center, SE-10691 Stockholm, Sweden}

\author{F. A. G\'omez Albarrac\'in}
\email{albarrac@fisica.unlp.edu.ar}
\affiliation{Instituto de F\'isica de L\'iquidos y Sistemas Biol\'ogicos (IFLYSIB), UNLP-CONICET, Facultad de Ciencias Exactas, 1900 La Plata, Argentina}

\affiliation{Departamento de Ciencias B\'asicas, Facultad de Ingenier\'ia, Universidad Nacional de La Plata, 1900 La Plata, Argentina}

\author{M. \v{Z}ukovi\v{c}}
\email{milan.zukovic@upjs.sk}
\affiliation{Department of Theoretical Physics and Astrophysics, Institute of Physics, Faculty of Science, Pavol Jozef \v{S}af\'arik University in Ko\v{s}ice, Park Angelinum 9, 041 54 Ko\v{s}ice, Slovak Republic}

\author{H. D. Rosales}
\email{rosales@fisica.unlp.edu.ar}
\thanks{corresponding author}
\affiliation{Instituto de F\'isica de L\'iquidos y Sistemas Biol\'ogicos (IFLYSIB), UNLP-CONICET, Facultad de Ciencias Exactas, 1900 La Plata, Argentina}

\affiliation{Departamento de Ciencias B\'asicas, Facultad de Ingenier\'ia, Universidad Nacional de La Plata, 1900 La Plata, Argentina}

\date{\today}

\begin{abstract}
Skyrmions -topologically nontrivial magnetic quasi-particles- can emerge in two-dimensional chiral magnets due to moderate or high strength of the Dzyaloshinskii-Moriya (DM) interaction. In this work, we show that the inclusion of weak next-nearest-neighbor DM interaction in the frustrated $J_1$-$J_2$ honeycomb-lattice Heisenberg antiferromagnet leads to the emergence of field-induced incommensurate antiferromagnetic meron-antimeron pairs and antiferromagnetic skyrmion structures. Using the Luttinger-Tisza approximation and large-scale Monte Carlo simulations, we report that for lower frustration values, antiferromagnetic meron-antimeron pair crystal and gas phases emerge within a small magnetic field window. In these meron phases, the fundamental unit consists of a meron-antimeron pair residing on different sublattices of the honeycomb lattice, even in the gas phase, where they exhibit greater mobility. For larger frustration, a two-layer-like three-sublattice antiferromagnetic skyrmion crystal phase is stabilized over a wider magnetic field range. At lower temperatures, this region splits into two distinct antiferromagnetic skyrmion phases with skyrmions of different sizes, reflecting the influence of frustration and thermal effects on the stabilization of these topological textures. Interestingly, both meron and skyrmion low-temperature phases connect to spiral spin liquid phases of the honeycomb lattice at higher temperatures. Additionally, we analyze the emergent single-$q$ and double-$q$ phases at low temperatures, constructing a comprehensive phase diagram.

\end{abstract}

\maketitle

\section{Introduction}

The exploration of advanced spintronics has surged with the identification of various magnetic topological structures showing unconventional properties \cite{BeyondSky}. In particular, more recently antiferromagnetic spintronics \cite{gomonay2017concepts,baltz2018antiferromagnetic} stands at the forefront of this field, offering and exploring materials with unique characteristics that present a wealth of potential applications. 

A prominent example of a topological texture is the magnetic skyrmion \cite{Bogdanov1989,Bogdanov1994,Bogdanov2006}, a swirling spin texture that has garnered significant attention since its experimental observation in chiral magnetic materials \cite{MnSi,Yu2010,tonomura2012real,tonomura2012real,zhang2023material}. Skyrmions exhibit remarkable stability \cite{NagaosaTech,hagemeister2018controlled}, can be manipulated with low current densities \cite{NagaosaTech,lemesh2018current,peng2021dynamic}, and possess nanoscale dimensions, making them promising candidates for future spintronic devices.

These textures, exhibiting non-trivial topological properties, are characterized by a topological invariant known as the topological charge, defined as

\begin{equation}
    n_{sk}=\frac{1}{4\pi}\int \Sp\cdot(\partial_x\Sp\times \partial_y\Sp)d^2\rv 
\end{equation}
where $\Sp$ is the local unit vector field associated to the three-dimensional spin texture vector at site $\rv$. This quantity counts how many times $\Sp$ wraps the unit sphere \cite{NagaosaTech} and is further linked to the polarity and vorticity of the skyrmion, given rise to the emergence of skyrmions or antiskyrmions, depending on the underlying magnetic interactions and lattice symmetries\cite{nayak2017ASky,koshibae2016theory}.  
In systems lacking inversion symmetry, the Dzyaloshinskii-Moriya interaction, arising from relativistic spin-orbit coupling, plays a crucial role in stabilizing periodic arrays of magnetic skyrmions \cite{bogdanov2001,binz2006,Bogdanov2006}. The competition between ferromagnetic exchange and DM interaction induces these helical spin structures. Conversely, in centrosymmetric systems, both skyrmions and antiskyrmions can coexist in a degenerate state under competing exchange interactions \cite{Okubo2012,mohylna2022spontaneous}. Factors such as anisotropy \cite{Leonov2015}, anisotropic exchange or dipolar interactions can lift this degeneracy, favoring one type of texture over the other. 

Beyond skyrmions, a variety of other topological objects can emerge in magnetic materials \cite{BeyondSky}, each conferring unconventional physical properties to the system. Among them, antiferromagnetic skyrmions, present unique behavior, such as straight-line motion under current without a skyrmion Hall effect; vortices, merons \cite{kamiya2014magnetic,lin2015skyrmion,Yu2018,Wang2021,hayami2021,hayami2024}, fractional skyrmions\cite{Gao2020,Rosales2022,jena2022observation},   ferrimagnetic skyrmions \cite{SkyFerri}, antiferromagnetic skyrmions (and their antiskyrmion counterparts)\cite{Rosales2015,Osorio1,Fang2021,Zukovic1,villalba2019field,Zukovic2,Zukovic3,mohylna2022spontaneous,Pham2024,AFSky1,AFSky2}, antiferromagnetic skyrmion bubbles \cite{SkyAFBubbles}, biskyrmions\cite{yu2014biskyrmion}, spacetime topology with hopfions \cite{kent2021Hopfions,knapman2024spacetime}, and Bloch points\cite{lang2023bloch}. 

The stabilization and control of such intricate topological structures often rely on the underlying magnetic interactions and lattice geometries. In particular, magnetic frustration -a phenomenon arising from competing interactions or specific lattice geometries- plays a pivotal role in inducing novel magnetic states. This intrinsic competition among interactions has been shown to be instrumental in the formation of various topological textures, including ferromagnetic skyrmions \cite{Okubo2012,kawamura2024skyrmion}, antiferromagnetic skyrmions \cite{mohylna2022spontaneous}, high-topological-number skyrmions \cite{hu2024skyrmions}, and multiple-$q$ states \cite{Shimokawa2019}. By leveraging frustration, it is possible to enhance or control these topological states, offering promising pathways for innovative applications in spintronics.

A prototypical example of competing interactions is the antiferromagnetic $J_1$-$J_2$ spin model on a square lattice, which has been extensively studied \cite{Chandra1988,schmidt2007square,drisko2017topological}. Another notable example of great interest is the honeycomb $J_1$-$J_2$ spin model, which has also been thoroughly explored, revealing a wide range of fascinating phenomena \cite{Mulder2010,Shimokawa2019,huang2022spiral,khatua2024spin}. 
In the latter case, a subset of ground-state configurations forms a continuous manifold in reciprocal space, leading to the emergence of a spiral spin liquid (SSL) phase\cite{yan2022low}(a type of classical spin liquid characterized by a subextensive degeneracy of ground states). In this  regime, spins fluctuate around spiral contours in momentum space. This SSL phase has been recently observed experimentally in van der Waals (vdW) materials \cite{gao2022spiral,gao2024codimension}.

In this work, we explore the rich interplay between skyrmion physics and spiral spin liquid behavior in the frustrated $J_1$-$J_2$ honeycomb-lattice Heisenberg antiferromagnet, augmented by a very weak next-nearest-neighbor Dzyaloshinskii-Moriya interaction. This lattice geometry, characterized by its unique two-sublattice structure and small coordination number, offers a fertile ground for the emergence of unconventional magnetic phases driven by frustration.

Using the Luttinger-Tisza approximation (LTA) and large-scale Monte Carlo (MC) simulations, we uncover two distinct field-induced topological textures. At lower frustration values, we identify an antiferromagnetic meron-antimeron lattice phase, where the fundamental building block is a bound meron-antimeron pair residing on different sublattices of the honeycomb lattice. Remarkably, this phase may evolve into a gas-like state of mobile meron-antimeron pairs within a narrow magnetic field window. For higher frustration values, we find that each triangular sublattice of the honeycomb lattice hosts an antiferromagnetic skyrmion crystal, stabilized across a broader magnetic field range. At low temperatures, this region separates into two phases with skyrmions of distinct sizes, emphasizing the intricate role of frustration and thermal fluctuations in shaping the magnetic landscape. 

Importantly, both meron-antimeron and skyrmion phases exhibit a direct connection at high temperatures to the spiral spin liquid regime characteristic of the honeycomb $J_1$-$J_2$ model. This highlights the dual influence of lattice geometry and competing interactions in stabilizing exotic topological states. By combining analytical insights from LTA with thermal effects captured through MC simulations, we construct a comprehensive phase diagram, clarifying the interplay of frustration, chirality, and topology in honeycomb magnets.

The article is structured as follows. In Section \ref{sec:2}, we introduce the model and present a detailed analysis using the LTA at zero temperature and zero magnetic field. Here, we explore the phase diagram and discuss how small Dzyaloshinskii-Moriya interactions can lift the degeneracy of the ground state. In Section \ref{sec:3}, we focus on the finite temperature behavior, where we describe the results from Monte Carlo simulations, comparing them with the LTA predictions, and investigate the emergence of antiferromagnetic meron-antimeron lattices, antiferromagnetic skyrmion lattices, and both single-$q$ and double-$q$ phases. Section \ref{sec:4} provides analytical ans\"atze for describing the antiferromagnetic skyrmion and meron-antimeron textures in more detail. Finally, a summary of our findings and conclusions is given in Section \ref{sec:5}.

\begin{figure}[th]
\centering
    \includegraphics[width=0.8\columnwidth]{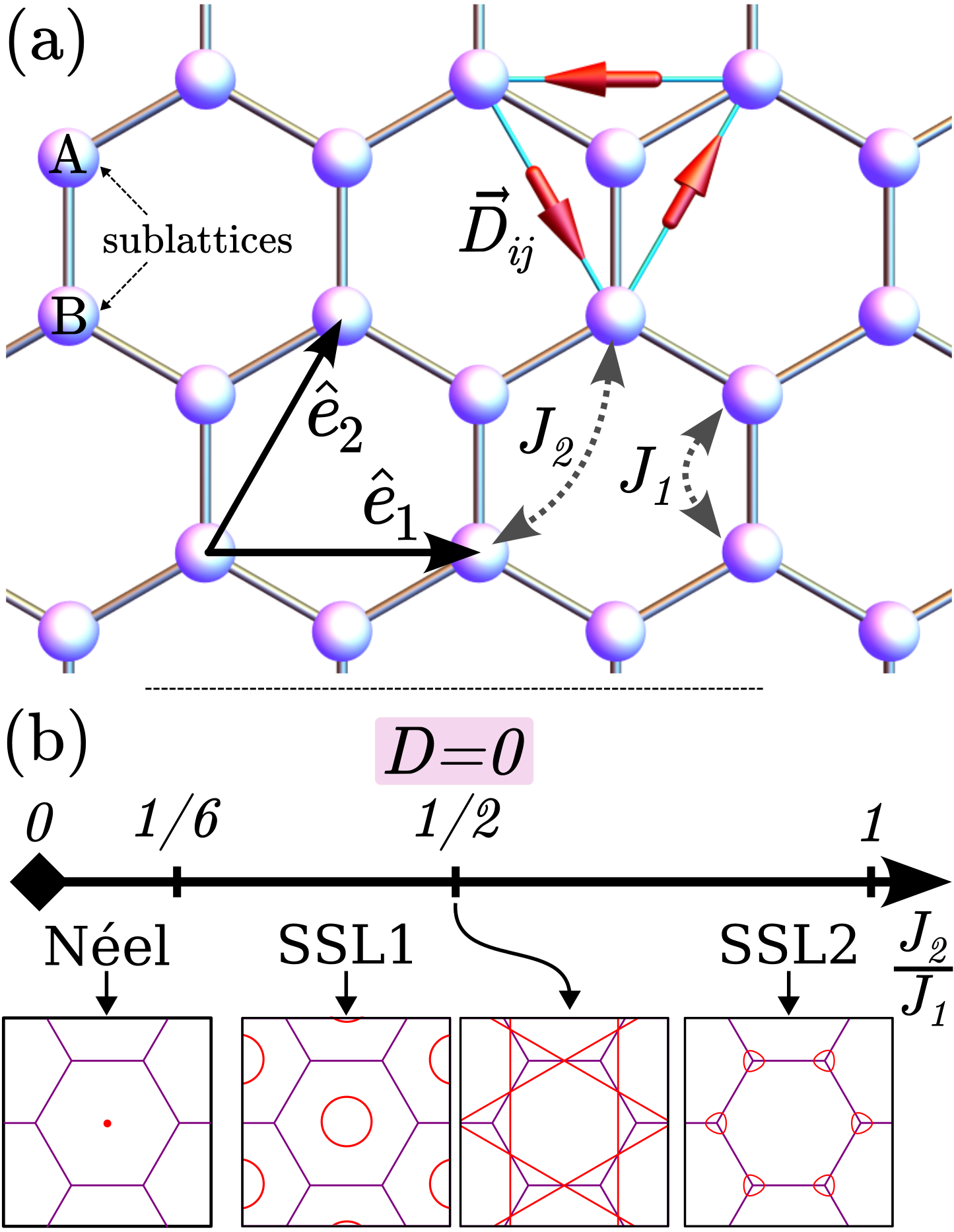}
    \caption{(a) Definition of the couplings $J_1,J_2$ and $\mathbf{D_{ij}}$, and basis vectors $\hat{e}_1,\hat{e}_2$ on the honeycomb lattice. (b) Ground state solutions at $T=0$ for the pure antiferromagnetic $J_1$-$J_2$ model ($D=0$) as a function of the $J_2/J_1$ ratio and examples in momentum space, showing the presence of:  a Néel state for $J_2/J_1 < 1/6$ at the $\Gamma$ point (red dot); two non-equivalent degenerate spiral spin liquid states, SSL1 ($1/6 < J_2/J_1 < 1/2$) and SSL2 ($J_2/J_1 > 1/2$), where the ground state exhibits a ring-like degeneracy around the $\Gamma$ and $K$ points, respectively (red lines). At the special point $J_2/J_1 = 1/2$, both spiral spin liquid states merge.}
    \label{fig:lattice}
\end{figure}
%

\section{Model and Zero Temperature Phase Diagram}
\label{sec:2}
In this paper we focus on the frustrated $J_1$-$J_2$ classical antiferromagnetic Heisenberg model on the honeycomb lattice under a magnetic field perpendicular to the lattice plane ($z$ direction) and include the effects of explicit spin-rotation symmetry breaking by the presence of the antisymmetric in-plane Dzyaloshinskii-Moriya term: 
\begin{eqnarray} \label{eq:H}
\mathcal{H}&=&J_1\sum_{\langle i,j\rangle}\Sp_i\cdot\Sp_j+J_2\sum_{\langle\langle i,j\rangle\rangle}\Sp_i\cdot\Sp_j \nonumber \\&&+\sum_{\langle\langle i,j\rangle\rangle} {\bf D}_{ij}\cdot(\Sp_i\times\Sp_j)- B\sum_i\,S_i^z
\end{eqnarray}
\noindent where $\Sp_i$ are unit-vector classical Heisenberg spins, $\{J_1,J_2\}$ are antiferromagnetic exchange interactions ($J_{1,2}>0$) to nearest and next-nearest neighbor pairs (denoted in the sums by $\langle i,j \rangle$ and $\langle\langle i,j \rangle\rangle$ respectively), ${\bf D}_{ij}=D\,(\rv_j-\rv_i)/|(\rv_j-\rv_i)|$ is the next-nearest-neighbors  Dzyaloshinskii-Moriya interaction with the orientation defined by the crystal symmetries and $B$ is the magnitude of the external magnetic field in the $z$ direction.  Note that in this lattice the DM interaction is forbidden on nearest-neighbors bonds because an inversion center lies exactly in the middle of the bond. Therefore, the first possible DM coupling in the honeycomb lattice is between next-nearest-neighbors. As we discuss below, we will see that even a small DM strenght $D \ll  J_1$ plays a crucial role in the stabilization of exotic topological 
textures under an external magnetic field. In Fig.~\ref{fig:lattice} (panel (a)) we indicate the exchange and DM couplings in the honeycomb lattice and the basis vectors chosen in this work. 

\subsection{Luttinger-Tisza Approximation ($T=0, B=0$)}  
\label{subsec:LTA}

To explore the classical ground state (GS) phase diagram in the absence of a magnetic field at zero temperature, we resort to the Luttinger-Tisza approximation (also known as the spherical model)\cite{Luttinger1946,Luttinger1951}. Within this scheme, instead of imposing the local ``strong constraint'' $|\Sp_i|=1$, one imposes a global so-called ``weak constraint'' $\sum_i|\Sp_i|^2=N\,S^2$, where $N$ is the number of lattice sites. With this softer constraint, the model Hamiltonian in Eq. (\ref{eq:H}) can be diagonalized by a simple Fourier transformation as $S^{a}_{\qv}=\frac{1}{\sqrt{N}}\sum_j\,S^{a}_je^{\qv\cdot\rv_j}$, where $a=x,y,z$ indicates the
spin component and $\qv$ and $\rv_j$ denote the pseudo-momentum and position respectively. In the case of the two-sublattice honeycomb lattice, the Hamiltonian then becomes

\begin{eqnarray}
\mathcal{H}&=&\sum_{\qv}\Psi_{-\qv}\cdot {\bf M(\qv)}\cdot\Psi_{\qv}
\label{eq:HLTA}
\end{eqnarray}
\noindent where $\Psi_{\qv}=\{S^x_{\qv,A},S^y_{\qv,A},S^z_{\qv,A},S^x_{\qv,B},S^y_{\qv,B},S^z_{\qv,B}\}$, $A,B$ are sublattice indices (as indicated in Fig.~\ref{fig:lattice})  and  ${\bf M(k)}$ is a $6\times 6$ complex matrix given by

\begin{equation}
{\bf M}(\qv)=\left[ \begin{array}{cc}
m_{11} & m^*_{12} \\
m_{12} & m_{11}
\end{array} \right]
\label{eq:matrixK}
\end{equation}
where 
\begin{equation}
\small
m_{11}=\left[ \begin{array}{ccc}
g_2&0&i\,g^y_3 \\
0&g_2&-i\,g^x_3\\
-i\,g^y_3&i\,g^x_3&g_2
\end{array} \right],\,[m_{12}]_{ij}=g_1\,\delta_{ij},\nonumber
\end{equation}
\noindent$g_1=\frac{J_1}{2}\sum_{{\bf \delta_1}}e^{i\,\kv\cdot{\bf \delta_1}}$, $g_2=J_2\sum_{{\bf \delta_2}}\cos(\qv\cdot{\bf \delta_2})$ and $\vec{g}_3=\{g^x_3,g^y_3\}=D\sum_{{\bf \delta_2}}\sin(\qv\cdot{\bf \delta_2})\,{\bf \delta_2}$. Here ${\bf \delta_1}, {\bf \delta_2}$ indicate first and second neighbors, respectively with $\{{\bf \delta_1} \}\equiv\{\hat{0},\hat{e}_1,\hat{e}_2-\hat{e}_1\}$ and $\{{\bf \delta_2} \}\equiv\{\hat{e}_1,\hat{e}_2,\hat{e}_2-\hat{e}_1\}$ , where $\hat{e}_1=\hat{x}$ and $\hat{e}_2=\hat{x}/2+\sqrt{3}\hat{y}/2$ are unit vectors (see panel (a) in Fig.~\ref{fig:lattice}).


In this approach, the classical ground-state energy is determined by the minima of the lowest eigenvalue of the matrix ${\bf M(q)}$, which defines the ordering wave vector $\qv^*$ \cite{niggemann2019LTA}. For the model in Eqs.~(\ref{eq:H}) and (\ref{eq:HLTA}), it is straightforward to find that the lowest band is given by $w_{\text{min}}(\qv) = g_2 - |\vec{g}_3 + |g_1||$, and we will see that the ordering wave vectors $\qv^*$, as well as their number and position, reflect the underlying competition between interactions.

Let's begin the study by discussing the possible phases in the ground state of the model in Eq.~(\ref{eq:H}) for $B=0$ and $D=0$. The details of this model were thoroughly discussed in Refs.~\cite{Mulder2010,Shimokawa2019,huang2022spiral}, and the key results can be summarized as follows (see Fig.~\ref{fig:lattice} panel(b)):  for $J_2/J_1<1/6$ the GS exhibits N\'eel order with $\qv=0$. For $J_2/J_1>1/6$, it presents a family of degenerate incommensurate ring-like spin spiral ground states: for $1/6<J_2/J_1<1/2$ there is a single closed loop around $\qv=0$ in the first BZ; then at the special point $J_2/J_1=1/2$ marks where the contour touches the Brillouin zone boundary at the $M$ points. For $J_2 > 1/2$,  several spiral contours emerge around the $K$ points. This behavior closely resembles that observed in the $J_1-J_2-J_3$ antiferromagnetic triangular lattice along the special line $J_2 = 2\,J_3$, where degenerate momentum vectors also form spiral contours \cite{mohylna2022spontaneous}.

\begin{figure*}[th!]
\centering
    \includegraphics[width=0.85\textwidth]{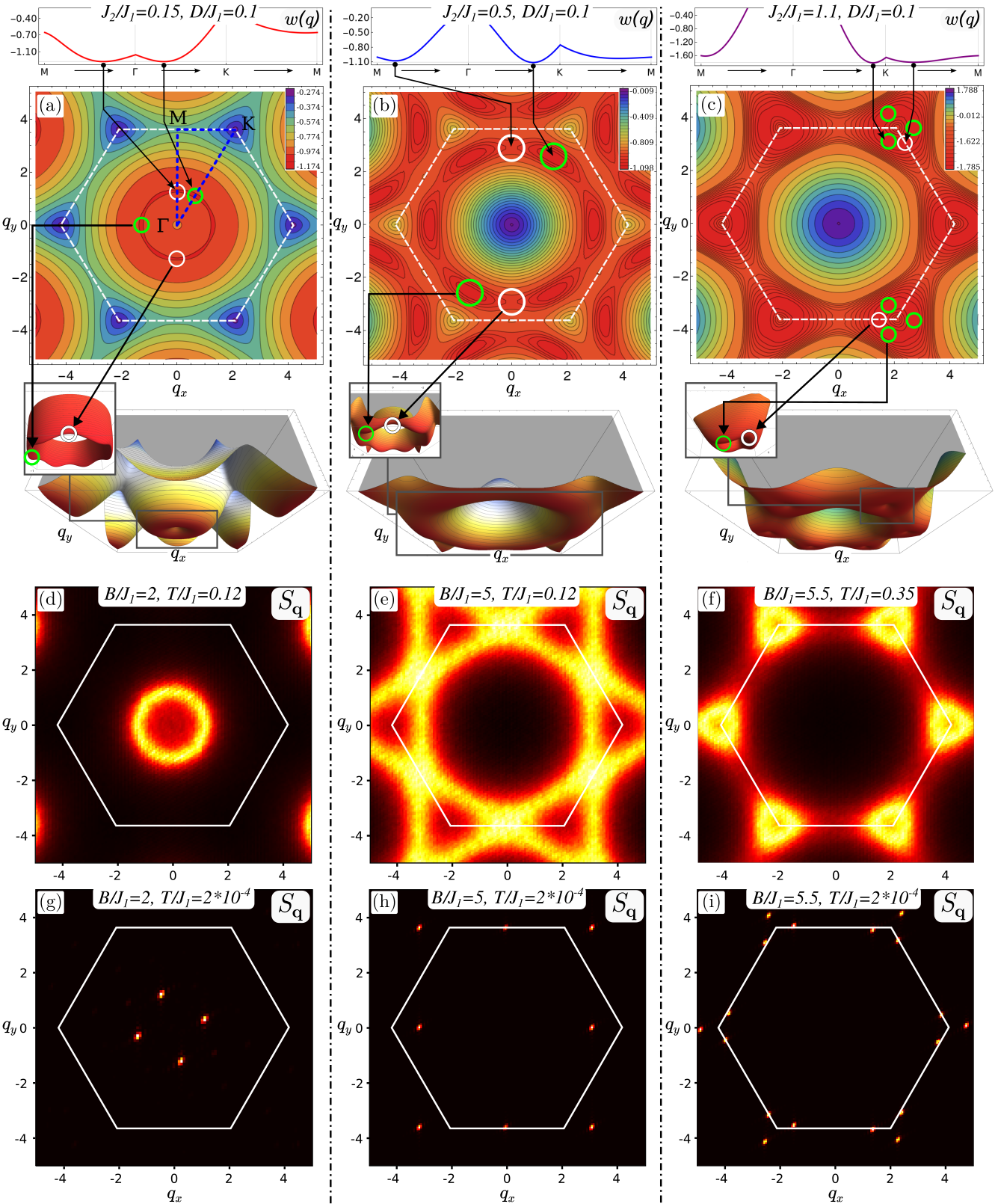}
    \caption{Top row:  LTA results for $D/J_1=0.1$ and $J_2/J_1=0.15,0.5,1.1$ (panels a-c). In each panel, at the center, the density plot of the lowest energy band indicates the absolute (green circles) and local (white circles) minima. On top, the detail of the minima along the $M \to \Gamma\to K\to M$ line. Below, is the 3D band structure.  Middle and bottom row: structure factors obtained from MC simulations for higher (d-f) and lower (g-i) temperatures for the same $J_2/J_1$ set at different values of the magnetic field. In panels (d-f), the characteristic SSL $S_\qv$ from the pure $J_1$-$J_2$ model in the honeycomb lattice is seen. At lower temperatures, different ordering vectors are selected among the degenerate semi-extensive high-temperature lines: for $J_2/J_1=0.15, B/J_1=2$ (panel g), the $S_\qv$ shows two inequivalent perpendicular ordering vectors, typical of meron lattices; for $J_2/J_1=0.5, B/J_1=5$ (panel h), there is a double-$q$ pattern, and for  $J_2/J_1=1.1, B/J_1=5.5$ a triple-$q$ structure with peaks in the borders of the Brillouin zone emerges.} 
    \label{fig:LTAlowband}
\end{figure*}

The system undergoes significant changes for $D>0$,  as this term favors a specific spin rotation direction, breaking the degeneracy observed at $D=0$.  For small but finite $D/J_1$,  the degenerate spiral contours may collapse into distinct points in momentum space, leading to a well-defined spin spiral ground state with a specific wave vector  $\qv$.  Our objective is to investigate the effects of a small $D/J_1$, under the hypothesis that the spiral spin liquid features of the pure Heisenberg model may still be accessible through thermal fluctuations. To define a baseline for a weak DM interaction, we fix the ratio of the DM strength to nearest-neighbor exchange interaction, $D/J_1=0.1$ throughout this study. Figure~\ref{fig:LTAlowband} shows a heat map of the energy dispersion of the lowest band, $w_{\text{min}}(\qv)$, for different values of the next-nearest-neighbor exchange interaction: $J_2/J_1=0.15$ (a), $J_2/J_1=0.5$ (b), and $J_2/J_1=1.1$ (c). The key features of the energy landscape are highlighted with green circles for the absolute minima and white circles for the local minima.

On top of these density plots, we present the energy dispersion $w_{\text{min}}(\qv)$ along the symmetry-allowed path $M \to \Gamma \to K \to M$, with the corresponding 3D structures of these bands displayed below (a-c). A key feature that emerges from these plots is the presence of three nonequivalent energy minima along the $\Gamma\to K$ direction in all cases, which suggests the possibility of stabilizing triple-$q$ structures under the influence of thermal fluctuations and magnetic fields. This is particularly significant as it implies that the system can accommodate complex magnetic configurations that are not limited to a single wave vector, thereby enriching the range of possible topological phases. For $J_2/J_1=0.15 \sim 1/6$ (panel (a)), the small energy barrier around the absolute minima indicates that thermal excitations could lead to the formation of a ring-like pattern in momentum space, which suggests that the system may exhibit exotic behaviors, such as the formation of spiral-like textures. This is consistent with the fact that $J_2/J_1 = 1/6$ marks the critical point for semi-extensive degeneracy (SSL1) in the pure Heisenberg model. At $J_2/J_1 = 0.5$ (panel (b)), the energy barrier around the absolute minima increases, and a new local minimum along the $\Gamma\to M$ direction emerges, which is similar to the behavior seen for smaller $J_2/J_1$ values but with a more pronounced structure. This suggests that as frustration increases, the energy landscape becomes more stabilized, suppressing certain fluctuating modes. Finally, for $J_2/J_1 = 1.1$ (panel (c)), the energy barriers around the absolute minima remain low, but new local minima appear along the $K \to M$ direction near the edges of the Brillouin zone, highlighting the potential for additional magnetic phases to emerge at these points, further expanding the phase space.

\begin{figure}
    \centering
    \includegraphics[width=1.0\columnwidth]{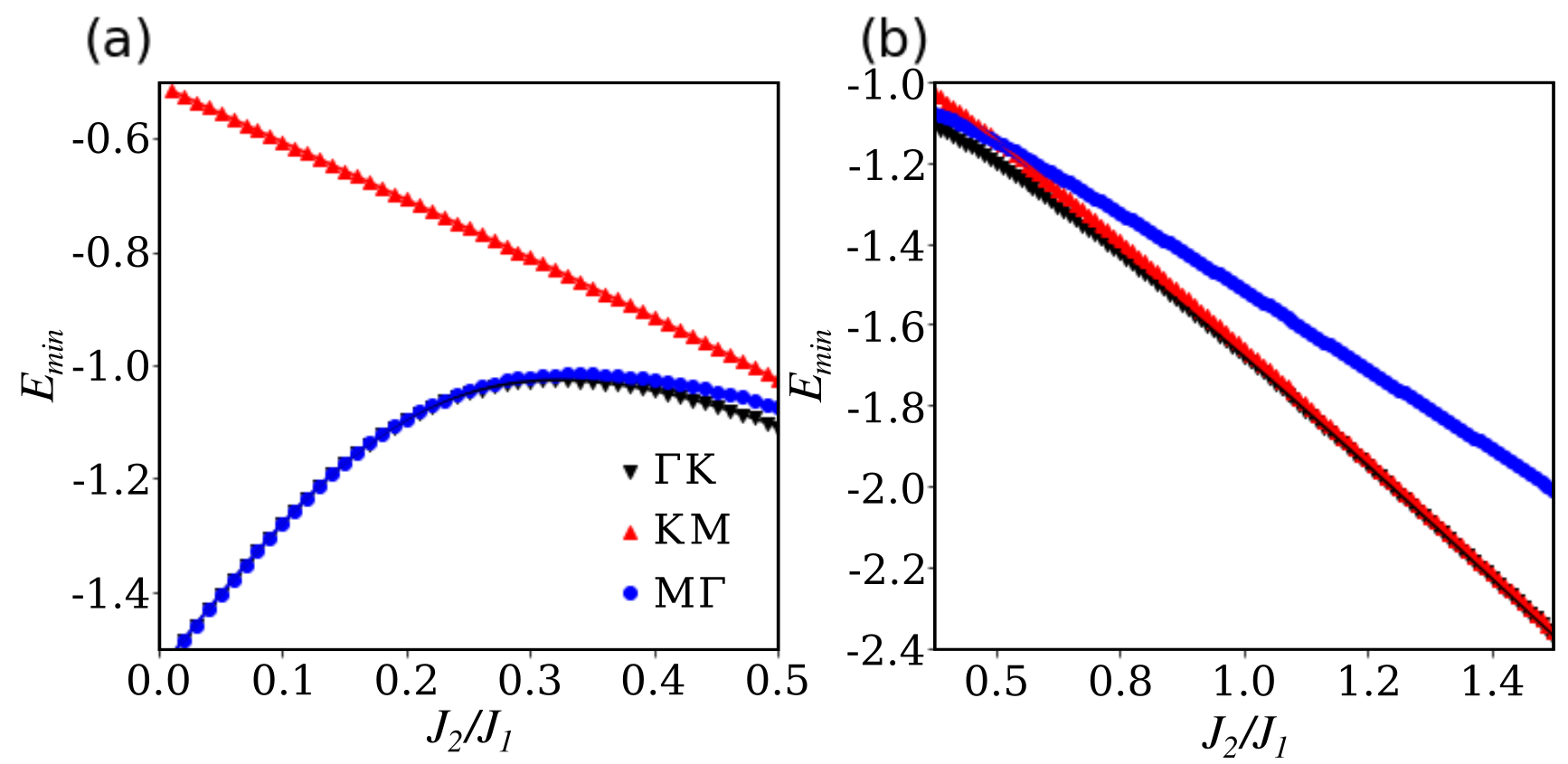}
    \caption{Comparison of the energy minima along three symmetry allowed paths in the BZ ($\Gamma \to K, K \to M, M \to \Gamma$) as a function of $J_2/J_1$ for $J_2/J_1<0.5$ (a) and $0.5<J_2/J_1<1.5$ (b), obtained from the LTA analysis for the $J_1$-$J_2$ honeycomb lattice with additional DM interaction of strength $D/J_1=0.1$. }
    \label{fig:LTAEmin}
\end{figure}

To further analyze these features, Fig.~\ref{fig:LTAEmin} compares the energy minima along three different symmetry paths: $\Gamma \to K$, $K\to M$, $M\to \Gamma$ as a function of $J_2$. Although the absolute minima consistently lie along the $\Gamma \to K$ line, the behavior driven by $J_2$ varies across different regimes: 

\begin{itemize}
\item{{\bf Lower} $J_2/J_1 \lesssim 0.3$: The energy minima along the $\Gamma\to K$ and $M\to \Gamma$ lines are very close, indicating that the degenerate ring found at $D=0$ is only slightly distorted. The energy barrier between the absolute minima and the rest of the ring is minimal, making ring-like states accessible through thermal fluctuations.}
\item{{\bf Intermediate} $0.3 \lesssim J_2/J_1 \lesssim 0.8$: The energy minima along the $K\to M$ path gradually approach those on the $\Gamma\to K$ line as $J_2$ increases, while the minima along the $\Gamma\to M$ path move further away.}
\item{{\bf High} $J_2/J_1 \gtrsim 0.8$: The $K\to M$ minima come significantly closer to the absolute minima along $\Gamma\to K$, which remains the global minimum for higher $J_2/J_1$. However, the small energy difference suggests that thermal fluctuations may stabilize states associated with the $K\to M$ minima.}
\end{itemize}

The preceding analysis indicates that, while the semi-extensive degeneracy of the pure Heisenberg model is lifted by a small $D$, the energy bands remain only slightly distorted from their $D=0$ counterparts. This subtle distortion suggests that other wave vectors $\qv$ may still become thermally accessible through fluctuations. These results highlight the interplay between weak Dzyaloshinskii-Moriya interactions and the inherent frustration of the $J_1$-$J_2$ honeycomb lattice, which may give rise to exotic magnetic behaviors at finite temperatures. 
To further investigate this possibility, we delve into the finite-temperature regime in the following section, employing large-scale Monte Carlo simulations to explore the combined effects of temperature and magnetic field on the stabilization of these states.

\begin{figure}[ht!]
    \centering
    \includegraphics[width=0.95\columnwidth]{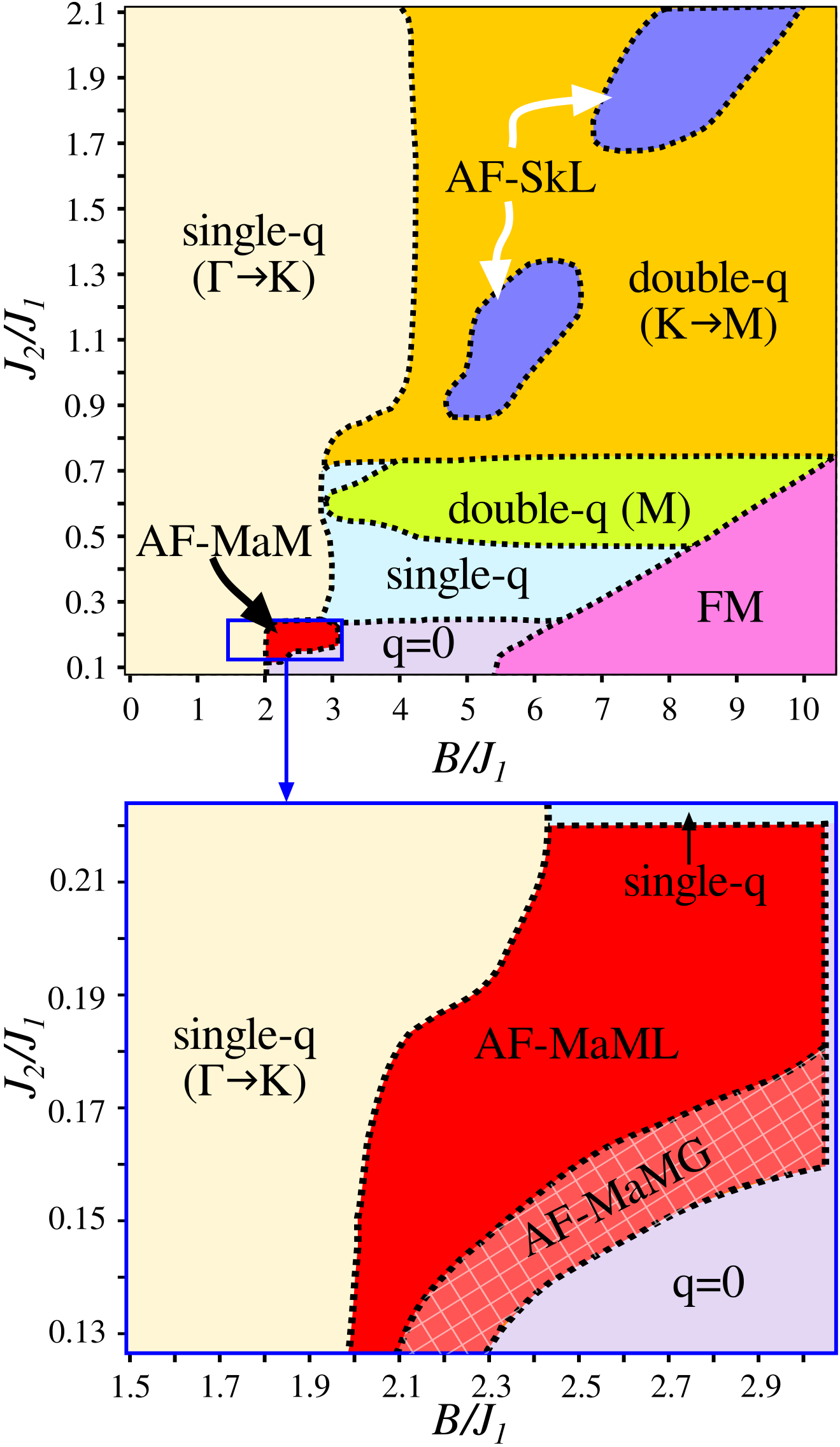}
    \caption{(Top) Low temperature ($T/J_1=2\times10^{-4}$) phase diagram as a function of frustrating coupling $J_2$ and external magnetic field $B$ obtained from Monte Carlo simulations for the Hamiltonian presented in Eq.(\ref{eq:H}) in the honeycomb lattice, fixing $D/J_1=0.1$. (Bottom) Zoom on the phase diagram presented in the top panel, focusing on the antiferromagnetic meron-antimeron region.}
    \label{fig:Diag}
\end{figure}
%

\section{Finite temperature behavior - Monte Carlo simulations}
\label{sec:3}

In the previous section, our Luttinger-Tisza analysis demonstrated that a weak DM interaction ($D\ll  J_1$) lifts the degeneracy of the pure Heisenberg model and enables the stabilization of multiple-$q$ textures. The small energy barriers between minima suggest that thermal fluctuations could promote the selection of different ordering patterns or even a partial recovery of spiral spin liquid (SSL) behavior. Previous studies on triangular-lattice-based systems have shown that the interplay between antiferromagnetic exchange interactions and DM interactions can lead to complex skyrmion textures, which emerge at finite temperatures under an external magnetic field \cite{Rosales2015,villalba2019field,Fang2021,Zukovic1,Rosales2022,Zukovic2,Zukovic3}.

In the honeycomb lattice model described by Eq.~(\ref{eq:H}), the strength of the frustrating exchange coupling, $J_2$, plays a crucial role in determining the energy band structure at $T=0$ and $B=0$. Given the bipartite nature of the honeycomb lattice, with two triangular sublattices coupled via $J_2$, and the fact that the DM interaction acts within each sublattice, it is expected that antiferromagnetic skyrmion lattices will stabilize when $D, J_2 \gg J_1$. Such lattices have been previously observed in triangular antiferromagnetic systems with in-plane DM couplings under a magnetic field \cite{Rosales2015,Fang2021,Zukovic1,Zukovic2,Zukovic3}.

However, our primary focus here is to determine whether a small DM interaction is sufficient to induce topological chiral textures in the frustrated $J_1$-$J_2$ honeycomb lattice. To address this, we fix $D/J_1=0.1$ and explore a wide range of values for the frustrating exchange coupling $0.1\leq J_2/J_1\leq 2$. Another crucial question is whether SSL features, characteristic of the pure Heisenberg model, can reemerge at finite temperatures through thermal fluctuations. 

To tackle these points, we perform extensive Monte Carlo simulations incorporating annealing, parallel tempering \cite{Swendsen1986} (PT), and over-relaxation \cite{Creutz1987} (OR) techniques. Simulations are carried out for system sizes $N=2\times N_c$ ($N_c=L^2$ being the number of cells) with lattice size $L=24$ to $72$, covering a broad temperature and magnetic field range to systematically investigate the possible emergence of non-trivial chiral phases. We note that using more sophisticated MC simulation approaches, such as PT, is essential in the study of systems with a complex energy landscape, such as frustrated spin models. In particular, PT allows the system to overcome entropic barriers in the free energy landscape by simulating a broad range of temperatures and thus escape metastable states. In our PT simulations, we used $160–300$ replicas (temperatures), the distribution of which follows the geometric progression. Such a setup aims to improve the diffusion (swapping) of replicas particularly in the generally problematic low-temperature region and thus increase the probability of finding stable states. The PT simulations are implemented on general-purpose graphical processing units using CUDA. In each replica, we first perform $5–9 \times 10^6$ hybrid (Metropolis + OR) MC sweeps to bring the system to equilibrium, and then another $5–7 \times 10^6$ sweeps are used to calculate the mean values. Replica swapping is proposed after each hybrid sweep through the whole lattice.

For analyzing the magnetic phases, we choose to describe them in terms of the triangular sublattices that make up the honeycomb lattice. This approach is convenient due to the bipartite nature of the lattice \cite{Shimokawa2019}, as it allows us to characterize the emergent phases with well-defined quantities.

As the first quantity of interest, we introduce the scalar chirality $\chi$ defined by summing the contributions from all elementary triangles of the sublattice. This provides a discrete measure of the topological charge $n_{sk}$,

\begin{equation}
\chi=\frac{1}{4\pi}\sum_{i=1}^N \Sp_i\cdot\left( \Sp_j\times\Sp_k\right)
\end{equation}
\noindent where $i,j,k$ are three spins in a triangular plaquette. In the continuum limit, a skyrmion has topological charge $n_{sk}=-1$; in a lattice (depending on the size of the texture) the presence of a skyrmion would correspond to $\chi\sim-1$. Similarly, other spin textures, such as merons (with $n_{sk}=-1/2$, representing a half-sphere spin configuration), also produce non-zero chirality values \cite{lin2015skyrmion, Yu2018, Wang2021}.

The second quantity we evaluate is the structure factor, which is the Fourier transform of the spin configuration and can be directly compared with experimental results from neutron-scattering techniques. In the presence of a magnetic field $B$,  we focus on the planar components of the spins $(xy)$, perpendicular to $B$. The structure factor $S_{\qv}$ is calculated as:
\begin{equation}
S_\qv=\frac{1}{N_c}\sum_{a=x,y}\left\langle  \left(\sum_{\rv}S^a(\rv)e^{i\qv\cdot\rv}\right)^2 \right\rangle
\end{equation}
\noindent where the sum is over all $N_c$ cells within the sublattice.

We summarize our main findings in the low-temperature $J_2/J_1$ vs $B/J_1$ phase diagram presented in Fig.~\ref{fig:Diag}. The most remarkable outcome of this work is the identification of three distinct regions where two types of non-trivial topological phases emerge:

\begin{itemize} \item Antiferromagnetic Meron-Antimeron region (AF-MaM):
At low $J_2/J_1\sim 0.13-0.22$ and intermediate $B/J_1\sim 2-3$ , we observe an antiferromagnetic meron-antimeron lattice (AF-MaML). This phase consists of two interpenetrating periodic arrangements of meron-antimeron pairs, one in each triangular sublattice. At higher magnetic fields, this phase transitions into a small region of the antiferromagnetic meron-antimeron gas phase (AF-MaMG), where the bound meron-antimeron pairs remain intact but lose long-range positional order.

\item Antiferromagnetic Skyrmion Lattice (AF-SkL):  
For higher $J_2/J_1 > 0.9$, as frustration increases, we identify two regions where antiferromagnetic skyrmion lattices are stabilized. These phases are characterized by three interpenetrating skyrmion lattices, forming a complex antiferromagnetic skyrmion lattice structure in each triangular sublattice.  
\end{itemize}

Interestingly, stabilization of the meron-antimeron phases occurs for $0.13 \lesssim J_2/J_1 \lesssim 0.2$, near $J_2/J_1 = 1/6$. This critical value corresponds to the transition point in the pure $J_1$-$J_2$ antiferromagnetic Heisenberg model ($D=0$), where the Néel phase gives way to the spiral spin liquid (SSL1) phase. The latter is characterized by a semiextensive ring-like degeneracy around the $\Gamma$-points of the Brillouin zone, as discussed in the previous section.  

On the other hand, the skyrmion phases are stabilized for $J_2/J_1 > 0.9$, where the degeneracy in the pure Heisenberg model shifts to closed curves around the $K$-points, corresponding to the SSL2 phase. Between these two regions, the meron and skyrmion phases are separated by a single-$q$ and a double-$q$ phases at intermediate values $0.3 \lesssim J_2/J_1 \lesssim 0.45$ and $0.45 \lesssim J_2/J_1 \lesssim 0.7$ respectively. Notably, this includes $J_2/J_1 = 0.5$, where the critical transition between SSL1 and SSL2 occurs in the pure Heisenberg case.  

The influence of the spiral-spin-liquid behavior remains evident at higher temperatures across all these regions. This is illustrated in Fig.~\ref{fig:LTAlowband}, where we show the structure factor $S_\qv$ at both higher and lower temperatures for three representative points in the phase diagram:  
\begin{itemize}
\item $J_2/J_1 = 0.15, B/J_1 = 2$ (Fig.~\ref{fig:LTAlowband}, panels (d,g)),  
\item $J_2/J_1 = 0.5,\,\,\, B/J_1 = 5$ (Fig.~\ref{fig:LTAlowband}, panels (e,h)),  
\item $J_2/J_1 = 1.1,\,\,\, B/J_1 = 5.5$ (Fig.~\ref{fig:LTAlowband}, panels (f,i)).  
\end{itemize}
At higher temperatures (panels (d) and (f)), we observe the characteristic SSL1 and SSL2 patterns, while at lower temperatures, the systems order into distinct topological phases. For example, at $J_2/J_1 = 0.15$, the meron lattice is stabilized, whereas, at $J_2/J_1 = 1.1$, the skyrmion lattice dominates. Furthermore, for $J_2/J_1 = 0.5$ (panel (e)), the semiextensive degeneracy from the merging of SSL1 and SSL2 at higher temperatures gives way to a double-$q$ ordering at low temperatures.  

In the following subsections, we delve into the details of these multiple-$q$, non-trivial topological phases -antiferromagnetic merons and skyrmions- and briefly comment on the remaining single-$q$ and double-$q$ phases observed in Fig.~\ref{fig:Diag}, comparing the Monte Carlo results with the LTA analysis.  

\begin{figure*}[ht!]
    \centering
    \includegraphics[width=0.95\textwidth]{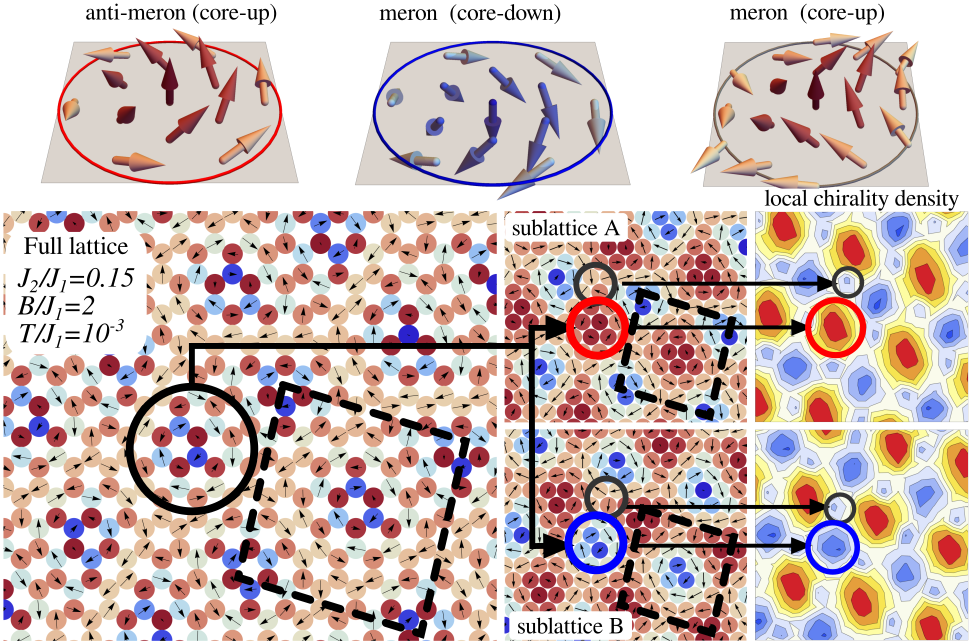}
    \caption{Details of the antiferromagnetic meron-antimeron lattice (AF-MaML), reminiscent of experimental results for Co$_8$Zn$_9$Mn$_3$ in Fig. 1 of Ref.\cite{Yu2018}. In the top row, the three types of merons realized in this texture are presented: core-up antimeron, core-down meron, and core-up meron. At the bottom, an example of low-temperature snapshot of the  AF-MaML order in the complete honeycomb lattice and of each triangular sublattice with its corresponding local chirality density (here $J_2/J_1=0.15, B/J_1=2,T/J_1=10^{-3}$). Dashed black lines indicate the unit cell of square meron/antimeron crystal.}
    \label{fig:MeronX}
\end{figure*}
\begin{figure}[h!]
    \centering
    \includegraphics[width=1.0\columnwidth]{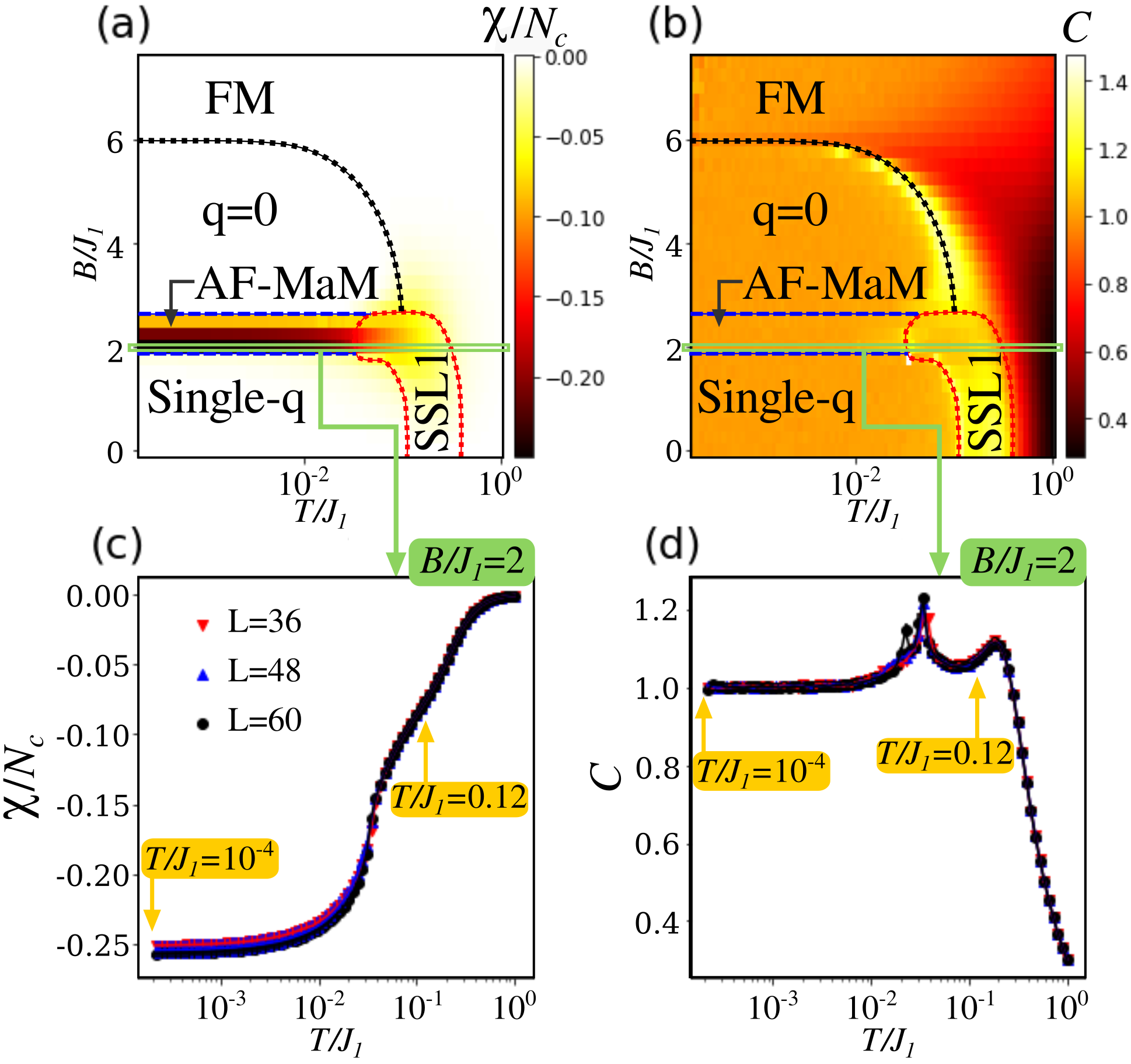}
    \caption{Chirality density (a) and specific heat (b) density plots as functions of temperature and magnetic field obtained from Monte Carlo simulations for $J_2/J_1=0.15$.  Panels (c) and (d) focus on the temperature dependence of the chirality density and specific heat, respectively, for $B/J_1=2$,  where the meron-antimeron crystal is stabilized, for $L=36, 48, 60$. Arrows indicate the temperatures where the $S_\qv$ are shown in Fig.~\ref{fig:LTAlowband}. Error bars, when not visible, are smaller than the size of the markers.} 
    \label{fig:ChiMeron}
\end{figure}
\begin{figure*}[ht!]
    \centering
    \includegraphics[width=0.9\textwidth]{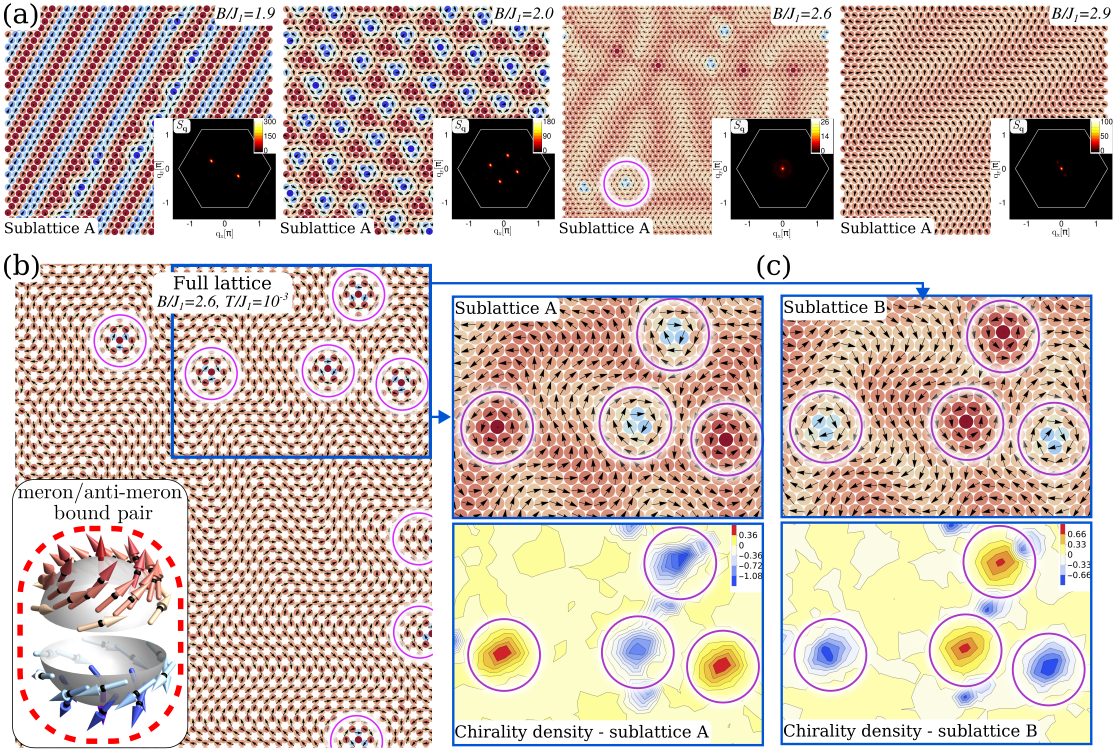}
    \caption{ (a) Example snapshots and their corresponding structure factors obtained from MC simulations for one triangular sublattice for $J_2/J_1=0.15$, depicting the low temperature  ($T/J_1=10^{-3}$) creation and annihilation of the AF-MaML phase with magnetic field, for  $B/J_1=1.9, 2, 2.6, 2.9$. At $B/J_1=2$ the meron-antimeron pairs form a periodic arrangement, or crystal, AF-MaML. At larger fields, $B/J_1=2.6$, the number of merons and antimerons is reduced, and they are not arranged periodically, hence the name meron-antimeron gas, AF-MaMG. (b,c) Detail of the AF-MaMG phase. (b) A region of the complete lattice at $B/J_1=2.6$, highlighting the meron-antimeron pairs between sublattices, illustrated in the inset. (c) Regions of each triangular sublattice their local chirality density are presented, showing that indeed the meron-antimeron pairs are bound between sublattices. }
    \label{fig:snapMAF}
\end{figure*}
%

\subsection{Antiferromagnetic meron-antimeron phases}

In this subsection, we provide a detailed analysis of the antiferromagnetic meron-antimeron (AF-MaM) phases that emerge at lower values of $J_2/J_1$ and $B/J_1 \sim 2$. To explore this behavior, we focus on the corresponding region of the phase diagram in the bottom panel of Fig.~\ref{fig:Diag}. As we discuss below, the formation of merons is preceded by a distortion of the single-$q$ phase, characterized by secondary peaks in reciprocal space equidistant from the $\Gamma$ point and perpendicular to the primary single-$q$ $\Gamma \to K$ pair. This distortion evolves into an antiferromagnetic meron-antimeron lattice (AF-MaML), consisting of two interpenetrated periodic arrangements of merons and antimerons, one in each triangular sublattice (Fig.~\ref{fig:MeronX}). 

In each sublattice, merons and antimerons organize into a square lattice, similar to the arrangement observed experimentally in Co$_8$Zn$_9$Mn$_3$ thin plates under an external magnetic field \cite{Yu2018}. As shown in Fig.~\ref{fig:MeronX}, this arrangement corresponds to a double-$q$ structure with orthogonal wavevectors, consistent with the spin textures imaged via Lorentz transmission electron microscopy in Ref.\cite{Yu2018}. Both merons and antimerons in our model share the same vorticity but differ in the orientation of their core relative to the magnetic field: merons have a core  aligned opposite to the field (core down, negative chirality), while antimerons have a core spin aligned along the field (core up, positive chirality). The maximum in-plane magnetization is found at the periphery of the merons and antimerons, while the cores exhibit nearly zero in-plane components, matching experimental observations. Between these textures, a smaller core-up meron is found, with vorticity $-1$ (see top row in Fig.~\ref{fig:MeronX}). 

Merons and antimerons with vorticity $+1$ form bound pairs between sublattices due to the antiferromagnetic coupling $J_2$ connecting the triangular sublattices. This is illustrated in Fig.~\ref{fig:MeronX}, where we show a detail of the full lattice, each sublattice, and its corresponding local chirality density. For each magnetic unit cell, indicated with dashed lines in Fig.~\ref{fig:MeronX}, the total $n_{sk}$ is $-1$ in both sublattices: there is one core-up antimeron with $n_{sk}=1/2$ in sublattice A (core-down meron with $n_{sk}=-1/2$ in sublattice B), surrounded by four quarters of core-down merons, $n_{sk}=4\times1/4\times-1/2$ (core-down antimerons $n_{sk}=4\times1/4\times1/2$ in sublattice B) and four halves of core-up merons in both sublattices ($n_{sk}=4\times1/2\times-1/2$), thus giving a total of $n_{sk}^A=1/2+4\times1/4\times-1/2+4\times1/2\times-1/2=-1$ for the magnetic unit cell in sublattice A and  $n_{sk}^B=-1/2+4\times1/4\times1/2+4\times1/2\times-1/2=-1$ in sublattice B. This non-zero $n_{sk}$ corresponds to a finite chirality density ($\chi/N_c$), a hallmark of topological order. 

To investigate the temperature-dependent behavior of the AF-MaML, we construct phase diagrams for the chirality density ($\chi/N_c$) and the specific heat $C=(\langle \mathcal{H}^2\rangle-\langle \mathcal{H}\rangle^2)/N\,T^2$  as functions of temperature and magnetic field, fixing $J_2/J_1 = 0.15$ (panels (a) and (b) in Fig.~\ref{fig:ChiMeron}). The $C$ exhibits a distinctive double-peak structure at lower magnetic fields,  which becomes more pronounced prior to the stabilization of the meron-antimeron phase for $B/J_1 \sim 1.9-2.6$. Focusing on $B/J_1 = 2$, we analyze $\chi/N_c$ and $C$ as functions of temperature for various system sizes (panels (c) and (d)). The chirality density converges to a constant value at the lowest temperature, while $C$ exhibits two notable features: a crossover at higher temperatures from the paramagnetic phase and a peak at $T/J_1 \sim 0.08$, indicating the stabilization of the AF-MaML phase. Between these transitions, a spiral spin liquid (SSL1) phase emerges, driven by thermal fluctuations.

The AF-MaML phase is further distinguished by its reciprocal space features. At higher temperatures, the system exhibits the semiextensive degeneracy of SSL1, characterized by a ring-like pattern around the $\Gamma$ point in $S_\qv$ (Fig.~\ref{fig:LTAlowband}d). At lower temperatures, this degeneracy gives way to the characteristic $S_\qv$ peaks of the meron-antimeron lattice (Fig.~\ref{fig:LTAlowband}g). The connection between the thermal fluctuation-induced degeneracy and the low-temperature order highlights the interplay of frustration and topology in the model.

A particularly intriguing feature of the phase diagram is the stabilization of a disordered antiferromagnetic meron-antimeron gas (AF-MaMG) at higher magnetic fields within $0.13 \lesssim J_2/J_1 \lesssim 0.18$. In this phase, the merons and antimerons are not periodically arranged, and their numbers in each sublattice are not necessarily equal. However, due to the antiferromagnetic coupling, each core-down meron in one sublattice remains bound to a core-up antimeron in the other. This gas phase, stabilized by an entropic effect through thermal fluctuations, is short-lived, vanishing under further increases in the magnetic field as the spins polarize into a simple $q=0$ structure.

The evolution of these phases with increasing magnetic field is illustrated in Fig.~\ref{fig:snapMAF}, which shows real-space snapshots and their corresponding $S_\qv$ patterns for $J_2/J_1 = 0.15$ and $B/J_1 = 1.9 - 2.9$ in panel (a). Panel (b) highlights the AF-MaMG phase, with meron-antimeron pairs circled, and a close-up of a bound pair in the inset. The local chirality density for each sublattice is shown in panel (c).

In summary, the antiferromagnetic meron-antimeron region presents a rich variety of textures, from a dense lattice to a diluted gas, with robust topological characteristics. In the next subsection, we delve into another prominent topological phase of this system: the antiferromagnetic skyrmion lattice.
\begin{figure*}[th!]
    \centering
    \includegraphics[width=0.9\textwidth]{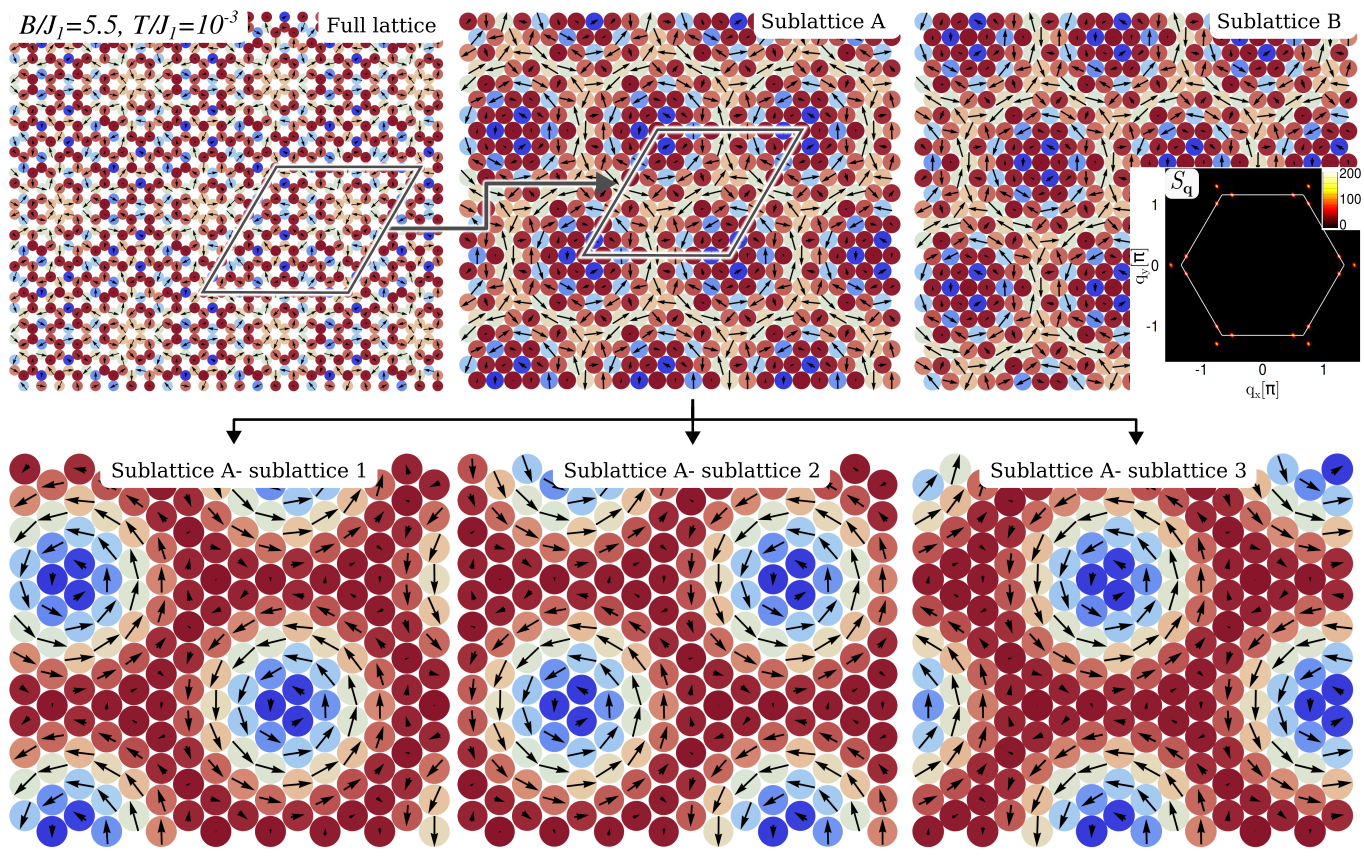}
    \caption{Detail of a typical case of AF-SkL phase obtained from MC simulations, for $J_2/J_1=1.1, B/J_1=5.5$ at $T/J_1=10^{-3}$, illustrating the real space snapshot for the total honeycomb lattice and the two triangular sublattices (top) and the three sublattices of one triangular sublattice (bottom). The structure factor of one triangular sublattice is shown, displaying triple-$q$ peaks in the BZ boundary. }
    \label{fig:snapAFSkX}
\end{figure*}
%

\subsection{Antiferromagnetic skyrmion phase}

As magnetic frustration increases and $J_2$ becomes comparable to or larger than $J_1$, two unconnected regions of skyrmion stability emerge in the low-temperature phase diagram (Fig.~\ref{fig:Diag}), showing that even with a small Dzyaloshinskii-Moriya interaction $D<J_1,J_2$, well-defined skyrmion structures can be stabilized in this system over significant areas of parameter space. Although separate, these two skyrmion regions share several common characteristics.  

First, unlike the antiferromagnetic meron-antimeron lattice (AF-MaML), the antiferromagnetic skyrmion lattice (AF-SkL) is stabilized within each triangular sublattice (i.e there is an AF-SkL on each sublattice (A,B) of the honeycomb lattice). Each AF-SkL consists of three interpenetrated $\sqrt{3}\times\sqrt{3}$ skyrmion lattices within the sub-sublattices ($1,2,3$), a feature reminiscent of previous findings in antiferromagnetic models on triangular lattices \cite{Rosales2015,mohylna2022spontaneous}. This structure is clearly illustrated by examining real-space and reciprocal-space configurations of the lattice, sublattice, and sub-sublattices, as shown in Fig.~\ref{fig:snapAFSkX}, obtained from a low-temperature Monte Carlo simulation for $J_2/J_1 = 1.1$ and $B/J_1 = 5.5$.  

\begin{figure}
    \centering
    \includegraphics[width=1.0\columnwidth]{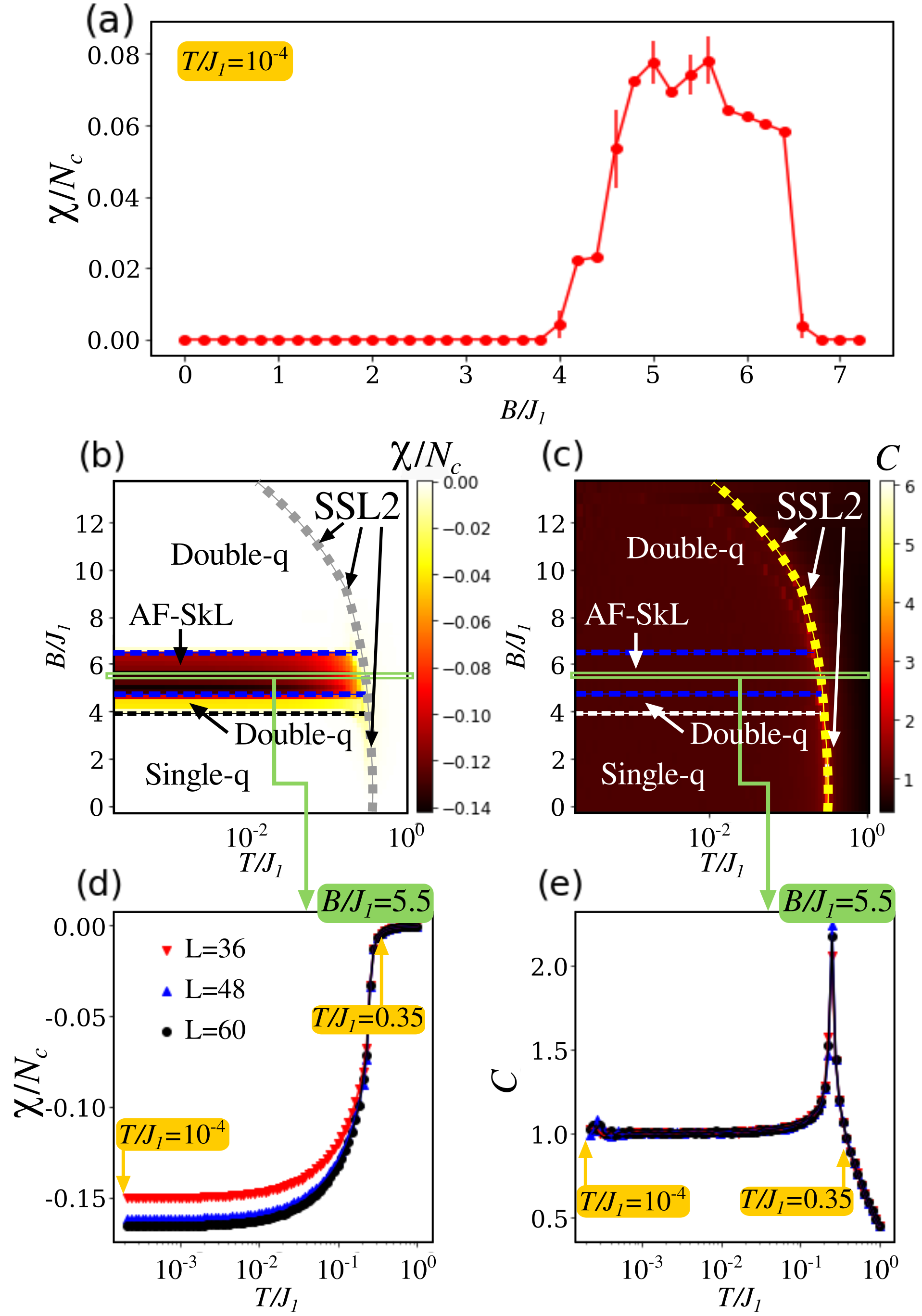}
    \caption{(a) Absolute value of the chirality density  as a function of magnetic field $B$ for $J_2=1.1$ at $T/J_1=2\times10^{-4}$ obtained from MC simulations, for $L=36$. Chirality scaled by $L^2$ (b) and specific heat (c) as a function of temperature for $J_2=1.1, B=5.5$ for system sizes $L=48,60,72$. Arrows indicate the temperatures where the $S_\qv$ are shown in Fig.~\ref{fig:LTAlowband}. When not seen, error bars are of the size of the markers or less.}
    \label{fig:ChisAFSky}
\end{figure}

The AF-SkL phase is stabilized across a broader magnetic field range, as confirmed by the calculation of the chirality density at low temperatures as a function of the field for $J_2/J_1 = 1.1$ (Fig.~\ref{fig:ChisAFSky}, panel (a)). Here, each point represents the averaged absolute value of the chirality over $10$ independent realizations for $L = 36$, with error bars indicating the standard deviation. In panels (b) and (c) we present the $\chi/N_c$ and $C$ density plots, showing the extension of the AF-SkL with temperature and magnetic field. Furthermore, the stability of the AF-SkL across different lattice sizes is confirmed in Fig.~\ref{fig:ChisAFSky}, panel (c), where the chirality density for three system sizes $(L = 48, 60, 72$) is plotted as a function of temperature for $J_2/J_1 = 1.1$ and $B/J_1 = 5.5$. The excellent agreement between system sizes highlights the robustness of the phase.

As anticipated, the AF-SkL order is evident in reciprocal space as a typical triple-$q$ pattern in the structure factor $S_\qv$. However, the position of these peaks is somewhat unconventional. While the $T = 0$ Luttinger-Tisza analysis predicted absolute minima along the $\Gamma\to K$ lines-fulfilling the $\sum_{i=1}^3 \qv_i = 0$ condition required for a skyrmion crystal \cite{BeyondSky}, Monte Carlo simulations reveal that thermal fluctuations drive the system into a triple-$q$ pattern where the $\qv$-peaks emerge along the $K \to M$ lines at the border of the Brillouin zone (BZ). These peaks correspond to secondary minima from the LTA along the $M\to\Gamma\to K\to M$ path, with only slight energy differences from the absolute minima.  

At higher temperatures, distinct peaks in $S_\qv$ vanish, giving way to degenerate lines around the $K$-points, reminiscent of the spiral spin liquid (SSL2) observed in the pure Heisenberg model (Fig.~\ref{fig:lattice}). This transition is driven by temperature-induced degeneracies, allowing the selection of other $\qv$-states as the system cools. Panels (f) and (i) in Fig.~\ref{fig:LTAlowband} illustrate the high- and low-temperature $S_\qv$ patterns for $J_2/J_1 = 1.1$ and $B/J_1 = 5.5$, showcasing the SSL2 and triple-$q$ configurations, respectively. 

Interestingly, the specific heat behavior as a function of temperature (Fig.~\ref{fig:ChisAFSky}, panel (c))  reveals a single transition from the paramagnetic phase to the low-temperature phases, preceded by the SSL2 regime. In panel (d) we focus on $C$ in the AF-SkL phase fixing $B/J_1=5.5$ for three different system sizes, showing a clear single peak. Dashed lines indicate the temperatures where example structure factors for the AF-SkL and the SSL2 phases are presented in panels (b) and (c) of Fig.~\ref{fig:ChisAFSky}. 

\begin{figure*}[th]
    \centering
    \includegraphics[width=0.9\textwidth]{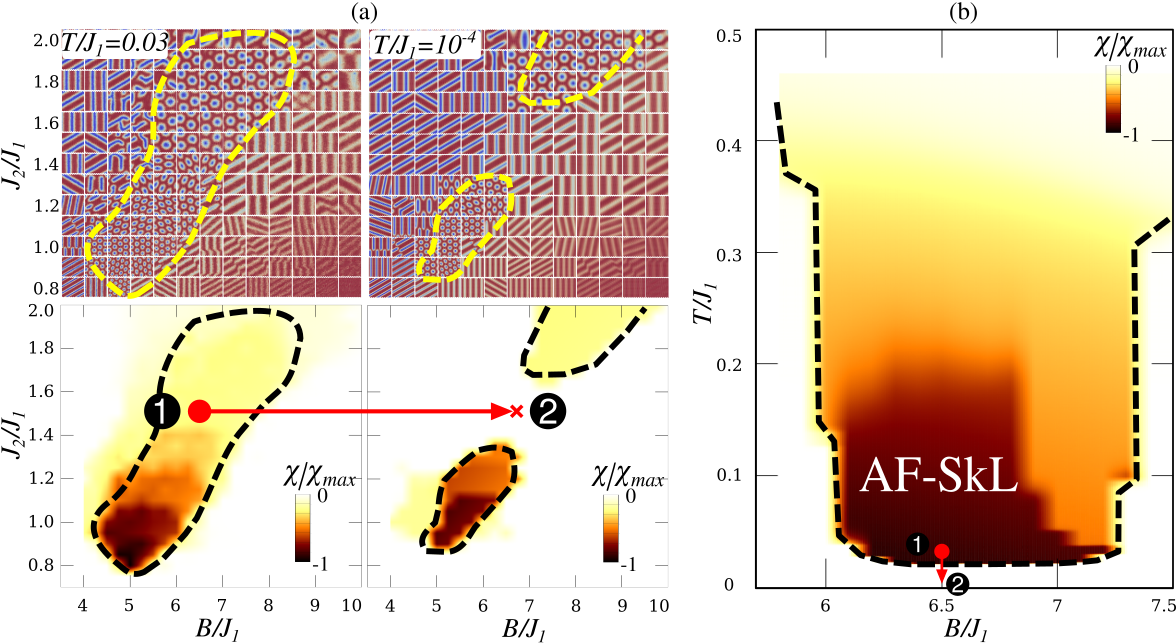}
    \caption{(a) Comparison of the scaled chirality density (bottom) and example snapshots (top) around the AF-SkL region as a function of $J_2/J_1$ and $B/J_1$ for two different temperatures, $T/J_1=0.03$ (left) and $T/J_1=10^{-4}$ (right). (b) Scaled chirality density plot as a function of temperature and magnetic field for $J_2/J_1=1.5$.}
    \label{fig:2TempsAFSkXl}
\end{figure*}

A notable feature of the low-temperature phase diagram (Fig.~\ref{fig:Diag}, top panel) is the separation between the two AF-SkL regions. Given their shared characteristics, it is reasonable to anticipate that these regions might merge under different external parameters. Indeed, at higher temperatures, this is observed: in the intermediate region of frustration $1.3 \lesssim J_2/J_1 < 1.7$, where double-$q$ states dominate at the lowest temperatures, the AF-SkL phase becomes stabilized over a wide magnetic field range for larger temperatures.  This behavior is demonstrated in Fig.~\ref{fig:2TempsAFSkXl}, panel (a). The bottom row compares the chirality density (scaled to the maximum absolute value, $\chi_{\text{max}}$) in the AF-SkL region at the lowest simulated temperature ($T/J_1 = 10^{-4}$) and a higher temperature ($T/J_1 = 0.03$) for $L = 36$. The top row provides snapshots of sub-sublattices at various phase diagram points, with dashed lines encircling the skyrmion regions.  From the snapshots, we deduce that stronger chirality at lower $J_2/J_1 \sim 0.8$ at $T/J_1 = 0.03$ corresponds to smaller skyrmions, leading to a higher skyrmion count $n_{sk}$. Conversely, as $J_2/J_1$ increases, skyrmions grow in size, reducing $n_{sk}$. Additionally, skyrmion crystals for larger $J_2/J_1 \gtrsim 1.9$ stabilize only at lower temperatures and vanish at higher $T/J_1$.  This temperature-dependent evolution of chirality is exemplified in panel (b) of Fig.~\ref{fig:2TempsAFSkXl}, which displays a scaled chirality density map as a function of temperature and magnetic field for $J_2/J_1 = 1.5$. At low temperatures, the system transitions sharply from the AF-SkL phase to a non-chiral double-$q$ state. The following subsection discusses this double-$q$ phase, its dominance at higher $J_2/J_1$, and the remaining single-$q$ and double-$q$ phases.  

\begin{figure}[ht!]
     \centering
     \includegraphics[width=0.99\columnwidth]{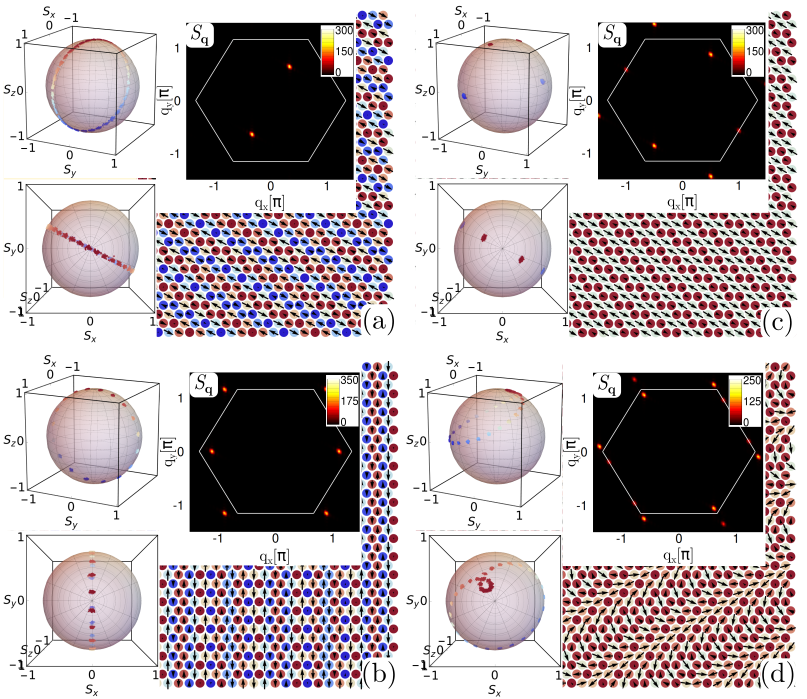}
     \caption{Example snapshots of one triangular sublattice with 3D projections of the spins and their corresponding structure factors for different single-$q$ and double-$q$ phases obtained at the lowest simulated temperature (a) single-$q$ phase for $J_2/J_1=0.3, B/J_1=0.5$ (b)  single-$q$ phase for $J_2/J_1=1, B/J_1=1$  (c) double-$q$ ($M$) phase for $J_2/J_1=0.6, B/J_1=4$ (d) double-$q$ ($K \to M$) phase for $J_2/J_1=1.2, B/J_1=8$}
     \label{fig:sqdq}
\end{figure}
%

\subsection{Single-$q$ and Double-$q$ phases}

Having described the topologically non-trivial phases that arise in this system, we now focus on the simpler single-$q$ and double-$q$ textures observed at low temperatures.

\subsubsection{Single-$q$ phase}
At low magnetic fields, for all values of $J_2$, the system stabilizes in a single-$q$ phase characterized by two peaks ($\qv$ and $-\qv$) along the $\Gamma\to K$ line in the Brillouin zone (BZ) of the triangular sublattices. These peaks represent a selection in each sublattice among the three single-$q$ pairs predicted by the LTA analysis. In real space, these textures manifest as helices where the projection along the magnetic field changes abruptly over short distances, due to the large value of $\qv$, which increases with $J_2$ and approaches the $K$ point.  

As the magnetic field is increased, these helical phases may become distorted, which is reflected in the structure factor by the appearance of additional peaks of much lower intensity, as shown in Fig.~\ref{fig:sqdq} (panels (a) and (b)). For lower values of $J_2/J_1 \lesssim 0.2$, the system transitions to a $q=0$ phase beyond the meron-antimeron region, as the magnetic field increases. For $0.2 \lesssim J_2/J_1 \lesssim 0.45$, distorted single-$q$ phases are stabilized before the system becomes fully magnetized.

\subsubsection{Double-$q$ phase}
Sharper and well-defined double-$q$ structures emerge for larger values of $J_2/J_1 \gtrsim 0.5$, particularly at higher magnetic fields. These double-$q$ phases are characterized by a second pair of bright peaks in addition to the $(\qv, -\qv)$ pair along the $\Gamma\to K$ line. Two distinct types of double-$q$ phases are observed:  

\begin{itemize}
    \item {\bf Double-$q$ ($M$)}: Found in a smaller region of the parameter space ($J_2/J_1 \sim 0.45 - 0.7$, $B/J_1 \gtrsim 3$), this phase is identified by a second pair of peaks at the $M$ points of the BZ. These peaks are located along a direction perpendicular to the $\Gamma\to K$ line (Fig.~\ref{fig:sqdq}c). The selection of these $M$-point peaks is influenced by the higher-temperature spiral spin liquid (SSL) behavior. Specifically, in this region, the higher-temperature $S_\qv$ pattern resembles the pure Heisenberg case for $J_2/J_1=0.5$, where the SSL1 and SSL2 degeneracy lines merge. This is evident when comparing the bottom panel of Fig.~\ref{fig:lattice} with panel (e) of Fig.~\ref{fig:LTAlowband}. In this SSL, the degeneracy lines are straight, connecting consecutive $M$ points in the BZ, making these points likely candidates for selection upon cooling.   
    \item {\bf Double-$q$ ($K\to M$)}: Dominating the low-temperature phase diagram at higher fields and larger $J_2/J_1 \gtrsim 0.8$, this phase features a second pair of peaks along the $K\to M$ line. These peaks correspond to the secondary minima from the LTA along the $M\to \Gamma \to K \to M$ path. This alignment mirrors the triple-$q$ peaks observed in the AF-SkL phase. In real space, the texture in this phase is a clear combination of helices that becomes increasingly polarized with the magnetic field, as illustrated in Fig.~\ref{fig:sqdq} (panel d).  
\end{itemize}
\section{Analytical ans\"atze for antiferromagnetic skyrmion and meron-antimeron textures}
\label{sec:4}

Following previous results obtained via the Luttinger-Tisza approximation and Monte Carlo simulations, the magnetic states emerging from the model described in Eq.~(\ref{eq:H}) can be effectively parametrized using a few dominant Fourier components. In this final section, we focus on deriving approximate analytical ans\"atze for the antiferromagnetic meron-antimeron and skyrmion lattice states.

To construct these ans\"atze, we expand the local magnetic moment as:
\begin{eqnarray}
S^{a}_j&=&\sum_{\nu=1}^{N_q}A^{a}_{\qv_\nu}\,e^{i\,\qv_\nu\cdot \rv_j}
\label{eq:SiAnasazt}
\end{eqnarray}
where $a=x,y,z$, $\{\qv_\nu\}$ ($\nu=1,..., N_q$) are the wave vectors corresponding to the peaks of the structure factor obtained from MC simulations, and $\{A^{a}_{\qv_\nu}\}$ are complex three-component vectors representing the Fourier amplitudes.
To determine these amplitudes, we compute the Fourier transform of the local spin configuration as $P^{a}_{\qv_{\mu}}=\sum_{j}e^{-i\,\qv_\mu\cdot \rv_j}\,S^{a}_j$. From this expression, the Fourier amplitudes $\{A^{a}_{\qv_\nu}\}$ can be directly obtained by solving the linear system:

\begin{eqnarray}
{\bf A}^{a}={\bf G}^{-1}\cdot {\bf P^{a}}
\label{eq:AqAnasazt}
\end{eqnarray}

where ${\bf A}^a=(A^{a}_{\qv_1},A^{a}_{\qv_2},...,A^{a}_{\qv_{N_q}})$, ${\bf P}^a=(P^{a}_{\qv_1},P^{a}_{\qv_2},...,P^{a}_{\qv_{N_q}})$ and $[{\bf G}]^{\mu\nu}=\sum_{j}e^{i\,(\qv_\mu-\qv_\nu)\cdot\rv_j}$. This approach provides a straightforward method to derive approximate parametrizations for both the meron-antimeron and skyrmion lattice states, allowing for a direct comparison with the numerical spin textures obtained from simulations.

\subsection{Meron-antiMeron lattice ansatz}
Building on the Fourier expansion method, we propose an explicit analytical ansatz for the meron-antimeron lattice phase. This ansatz captures the incommensurate double-$q$ structure characteristic of this phase, where the magnetic texture in each sublattice arises primarily from the dominant wave vectors $\qv_1$ and $\qv_2$ in reciprocal space. These vectors satisfy the conditions $|\qv_1|=|\qv_2|$ and $\qv_1 \perp \qv_2$, ensuring the orthogonality observed in the texture. The ansatz obtained is given by: 
\begin{eqnarray} \label{eq:meron}
m^{x,y}_{j}&=&I_{xy}\sum_{a=1}^{2}\,\sin(\qv_a\cdot\rv_j+\theta_a)\,\hat{e}^{x,y}_{a} \\
m^{z}_{j}&=&I_0+I_z\sum_{a=1}^{2}\,\cos(\qv_a\cdot\rv_j+\theta_a) \nonumber
\end{eqnarray}
where $\Sp_j={\bf m}_j/|{\bf m}_j|$, $\hat{e}_{a}$ are two orthonormal vectors perpendicular to $\qv_a$, and $I_{xy}$,$I_{z}$ and $I_{0}$ are parameters controlling the in-plane and out-of-plane components of the spin texture. 
These parameters are derived from the Fourier coefficients $A^{a}_{\qv_\nu}$ from Eq.~(\ref{eq:SiAnasazt}) and (\ref{eq:AqAnasazt}), which are numerically determined using the spin configurations $\{S^{a}_j\}$ from Monte Carlo simulations. The spin components are then expressed in terms of cosine and sine functions, incorporating internal phases $\theta_a$ and amplitudes 
 $I_{0}, I_{xy}$ and $I_{z}$. This reformulation allows for a more intuitive interpretation of the spin structure, where $I_0$ captures the homogeneous part of the magnetic configuration, $I_{xy}$ and $I_{z}$ represent the amplitudes in-plane and out-of-plane spin components, respectively, and $\theta_a$ captures the phase shifts associated with each wavevector $\qv_a$. Figure~\ref{fig:anzats}(a-b) shows a comparison between a low-temperature snapshot for one sublattice obtained for Monte Carlo simulations (panel (b)) for $J_2/J_1=0.15, B/J_1=2$ and  the proposed ansatz  (panel (a)), where the parameters from Eq.~(\ref{eq:meron}) are presented in  Table \ref{tab}. The modes included in the ansatz capture the key features of the spin texture, showing excellent agreement with the results from the simulations.

\begin{table}[h!]
\centering
 \begin{tabular}{||c c c||} 
 \hline
  & AF-MaML & AF-SkL  \\ [0.5ex] 
 \hline\hline
 $\qv_1:$& $-\lambda_1\{\frac{1}{4},\frac{\sqrt{15}}{4}\}$ & $\lambda_2\{1,0.107\}$ \\ 
 $\qv_2:$& $-\lambda_1\{\frac{\sqrt{15}}{4},-\frac{1}{4}\}$ & $\lambda_2\{-0.59,0.812\}$ \\ 
 $\qv_3:$& - & $\lambda_2\{-0.4,-0.92\}$\\ 
 $\hat{e}_1:$& $\{\frac{\sqrt{15}}{4},-\frac{1}{4}\}$ & $\{\sqrt{3}/2,1/2\}$ \\ 
 $\hat{e}_2:$ & $-\{\frac{1}{4},\frac{\sqrt{15}}{4}\}$ & $\{-\sqrt{3}/2,1/2\}$  \\
 $\hat{e}_3:$ & - & $\{0,-1\}$ \\
 $\{\theta_a\}:$ & $\{-\frac{\pi}{3},\frac{2\pi}{3}\}$& $\{-\frac{4\pi}{5},-\frac{4\pi}{5},\frac{3\pi}{5}\}$ \\
 $I_{xy}:$ &  0.36 & 0.28 \\
 $I_{z}:$ &  0.18 & 0.23 \\
 $I_{0}:$ &  0.14 & 0.2 \\ [1ex] 
 \hline
 \end{tabular}
 \caption{Parameter values used in the analytical ansatz to reproduce the spin textures for the antiferromagnetic meron-antimeron and skyrmion lattice phases. $\lambda_1=1.3$ (AF-MaM phase), $\lambda_2=3.92$ (AF-SkL phase)}
 \label{tab}
\end{table}

\subsection{AF-skyrmion crystal ansatz}
\begin{figure}[ht!]
     \centering
     \includegraphics[width=0.99\columnwidth]{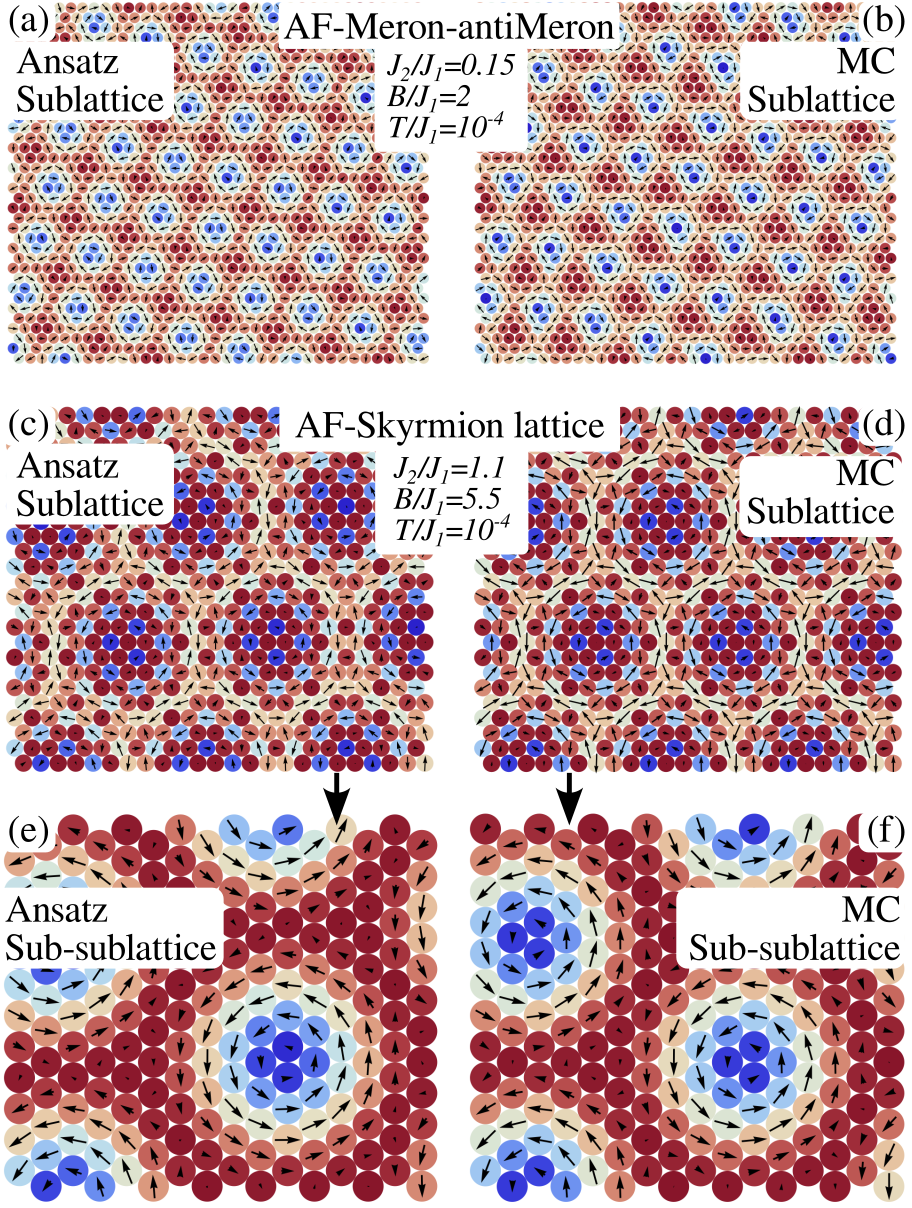}
     \caption{ Comparison of analytical ans\"atze and low-temperature snapshots obtained from simulations for triangular sublattices in the AF-MaML phase (a-b) and in the AF-SkL phase (c-d). (e-f) One sub-sublattice from panels (c-d).}
     \label{fig:anzats}
 \end{figure}

Similarly, for the AF-SkL phase, we propose a parametrization based on the typical superposition of three non-coplanar spirals within each triangular sublattice. Here, the ordering wave vectors $\qv_1,\qv_2,\qv_3$ satisfy the condition $\sum_{a=1}^3\qv_a=\pi$. Following the same approach as before, the ansatz is derived from the Fourier coefficients $A^{a}_{\qv_\nu}$, which are subsequently expressed in terms of amplitudes $I_{0}, I_{xy}$ and $I_{z}$ and phases $\theta_a$. The ansatz is given by:

\begin{eqnarray}
m^{x,y}_{j}&=&I_{xy}\sum_{a=1}^{3}\,\sin(\qv_a\cdot\rv_j+\theta_a)\,\hat{e}^{x,y}_{a} \\
m^{z}_{j}&=&I_0+I_z\sum_{a=1}^{3}\,\cos(\qv_a\cdot\rv_j+\theta_a) \nonumber
\end{eqnarray}
Notice that this set of equations is similar to that for the AF-MaML in Eq.~(\ref{eq:meron}), with the important difference that here the index $a$ runs over three $\qv$ vectors. Figure \ref{fig:anzats}(c-d) displays the magnetic structure of the AF-SkL phase as generated by the ansatz (panel (c)), with parameters from Table \ref{tab}, and compares it to the Monte Carlo results (panel (d)). We observe that the ansatz captures the essential features of the skyrmion lattice, including the three-sublattice periodicity (panel (e)) observed in the simulations (panel (f)).

%

\section{SUMMARY and CONCLUSIONS}
\label{sec:5}

In this work, we have explored the magnetic phases of the frustrated $J_1$-$J_2$ honeycomb-lattice Heisenberg antiferromagnet in the presence of a weak next-nearest-neighbor Dzyaloshinskii-Moriya interaction and an applied magnetic field. Through a combination of the Luttinger-Tisza approximation and large-scale Monte Carlo simulations, we constructed a comprehensive magnetic phase diagram, revealing a rich variety of topological and non-topological phases induced by frustration and field effects.

At moderate frustration values, we observed the stabilization of field-induced antiferromagnetic meron-antimeron pair lattice (AF-MaML) and gas (AF-MaMG) phases within a narrow magnetic field range. These phases feature antiferromagnetic meron-antimeron pairs that reside on different sublattices of the honeycomb lattice, with their mobility increasing in the diluted AF-MaMG phase. For higher frustration levels, the interplay between the DM and  $J_2$ interactions leads to the stabilization of two distinct antiferromagnetic skyrmion lattice phases (AF-SkL), characterized by skyrmions of different sizes. These skyrmion phases arise over a broader magnetic field range and exhibit transitions at low temperatures driven by thermal fluctuations and frustration.

We also uncovered a connection between these topologically non-trivial phases and spiral spin liquid (SSL) states at higher temperatures, highlighting the critical role of thermal fluctuations in the stabilization of emergent topological textures. Furthermore, at low temperatures and specific parameter regions, we analyzed simpler single-$q$ and double-$q$ magnetic structures. These phases, although topologically trivial, provide insight into the system's magnetic ordering mechanisms and its response to frustration and external fields.

Finally, we discuss the potential implications of our results for experimental studies on two-dimensional chiral magnets with honeycomb lattice geometries. Promising candidate materials include  Bi$_3$Mn$4$O${12}$(NO$3$)\cite{matsuda2010disordered}, FeCl$_3$ \cite{gao2022spiral}, GdZnPO \cite{Wan2024} and  CaMn$_2$P$_2$  \cite{Vaknin2025}, which have been shown to exhibit spin-liquid-like behavior and, in some cases, field-induced antiferromagnetic ordering.  These materials offer promising platforms for probing the interplay between frustration, single-site anisotropy, DM interaction, and field-induced topological phases, such as the meron-antimeron and skyrmion structures predicted in this work.
Our study shows the capability of weak DM to induce exotic magnetic textures in frustrated systems, broadening the understanding of topological phases in a honeycomb-lattice antiferromagnet.

\section*{Acknowledgments} 

M. M. acknowledges the financial support from SeRC and the Knut and Alice Wallenberg Foundation Grant No. 2018.0060, Grant No. 2021.0246, and Grant No. 2022.0108. Some of the computations/data handling were enabled by resources provided by the National Academic Infrastructure for Supercomputing in Sweden (NAISS), partially funded by the Swedish Research Council through grant agreement No. 2022-06725. M. \v{Z}. is supported by the grants of the Slovak Research and Development Agency (Grant No. APVV-20-0150) and the Scientific Grant Agency of the Ministry of Education of Slovak Republic (Grant No. 1/0695/23). F. A. G. A. and H. D. R. are partially supported by CONICET (PIP 2021-112200200101480CO),  SECyT UNLP PI+D  X947 and  Agencia I+D+i (PICT-2020-SERIE A-03205). F. A. G. A. acknowledges support from PIBAA 2872021010 0698CO (CONICET).


%

\end{document}